\pdfoutput=1

\documentclass[11pt]{article}

\usepackage[preprint]{acl}

\usepackage{times}
\usepackage{latexsym}

\usepackage[T1]{fontenc}

\usepackage[utf8]{inputenc}

\usepackage{microtype}

\usepackage{inconsolata}

\usepackage{graphicx}

\usepackage{xltabular}  
\usepackage{booktabs}   
\usepackage{array}      
\usepackage{ragged2e}   
\usepackage{longtable}
\usepackage{xtab}
\usepackage{makecell}
\usepackage{amsmath}

\newcommand{\eg}{\textit{e.g.}}

\newcommand{\ie}{\textit{i.e.}}

\title{A Survey of LLM-based Deep Search Agents: Paradigm, Optimization, Evaluation, and Challenges}

\author{
 \textbf{Yunjia Xi\textsuperscript{1}},
 \textbf{Jianghao Lin\thanks{Corresponding author}\textsuperscript{1}},
 \textbf{Yongzhao Xiao\textsuperscript{1}},
 \textbf{Zheli Zhou\textsuperscript{1}},
 \textbf{Rong Shan\textsuperscript{1}},
 \\
 \textbf{Te Gao\textsuperscript{2}},
 \textbf{Jiachen Zhu\textsuperscript{1}},
 \textbf{Weiwen Liu\textsuperscript{1}},
 \textbf{Yong Yu\textsuperscript{1}},
 \textbf{Weinan Zhang\textsuperscript{1}}
\\
\textsuperscript{1}Shanghai Jiao Tong University,
\textsuperscript{2}Central South University
\\
\texttt{\{xiyunjia,linjianghao,wnzhang\}@sjtu.edu.cn}
 }

\begin{document}
\maketitle
\begin{abstract}
The advent of Large Language Models (LLMs) has significantly revolutionized web search. The emergence of \textbf{LLM-based Search Agents} marks a pivotal shift towards deeper, dynamic, autonomous information seeking. These agents can comprehend user intentions and environmental context and execute multi-turn retrieval with dynamic planning, extending search capabilities far beyond the web. Leading examples like OpenAI's Deep Research highlight their potential for deep information mining and real-world applications. 
This survey provides the first systematic analysis of search agents. 
We comprehensively analyze and categorize existing works from the perspectives of architecture, optimization, application, and evaluation, ultimately identifying critical open challenges and outlining promising future research directions in this rapidly evolving field.
Our repository is available on \url{https://github.com/YunjiaXi/Awesome-Search-Agent-Papers}. 
\end{abstract}

\section{Introduction}



\begin{figure}
    \centering
    \vspace{-15pt}
    \includegraphics[clip,trim=7mm 4mm 7mm 3mm,width=0.5\textwidth]{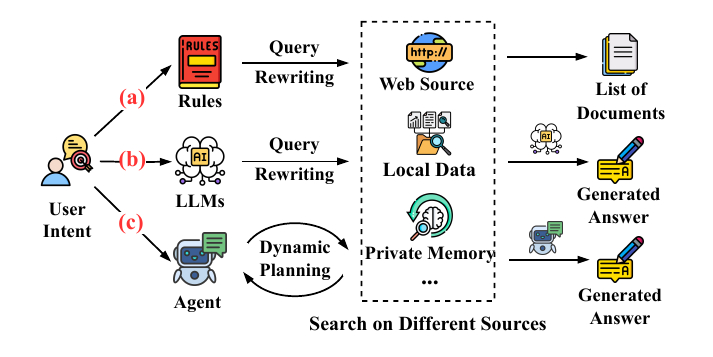}
    \vspace{-12pt}
    \caption{The evolution of search paradigm from (a) \textbf{Traditional Web Search} to (b) \textbf{LLM-enhanced Search}, and finally to (c) \textbf{Search Agents}.} 
    \vspace{-12pt}
    \label{fig:intro}
\end{figure}

The advent of Large Language Models (LLMs) has ushered in a new era of natural language processing, fundamentally transforming numerous fields, including web search~\citep{wang2024survey,zhao2023survey,hadi2023survey,xi2025bursting,lin2025can,lin2024rella,xi2025efficiency,xi2024towards}.
As shown in Figure~\ref{fig:intro}, \textbf{Traditional Web Search} required users to manually select and consolidate relevant information from a list of results~\citep{lin2021graph,dai2021adversarial,fu2023f}. With the rise of LLMs, \textbf{LLM-enhanced Search} emerged as a new paradigm, where LLMs rewrite user queries to improve search accuracy~\citep{ma2023query,liu2024query,xi2024memocrs} or summarize search results for quicker comprehension, \ie, traditional retrieval-augmented generation (RAG)~\citep{gao2023retrieval,fan2024survey}. However, this integration tends to be static, as LLMs rely on single-turn or rule-based iterative search, which struggles to handle complex and dynamic context effectively.

The emergence of LLM agents marked a pivotal shift, leading to \textbf{Search Agents}~\citep{zhang2024agentic}. Endowed with autonomy, search agents can control the entire search process, leveraging context more effectively for adaptive reasoning and dynamic retrieval. In this paradigm, search becomes a proactive action and is no longer limited to the web, but extends to a broader range of information sources, \eg, private databases and internal experiences within agents. Specifically, a search agent can be defined as \textit{an LLM agent capable of comprehending user intentions and environment contexts, autonomously planning search strategies, executing multi-turn dynamic retrieval from diverse sources, and integrating information to provide comprehensive insights}. Leading industrial solutions, \eg, Deep Research from OpenAI~\citep{openai_dr}, Gemini~\citep{gemini_dr}, and Perplexity~\citep{Perplexity_dr}, exemplify the potential of search agents in both deep information mining and commercialization.

Given these rapid advancements, we present the first systematic survey of search agents from multiple perspectives, analyzing them across the dimensions of \textit{how to search}, \textit{how to optimize}, \textit{how to apply}, and \textit{how to evaluate}. While recent related surveys have typically focused on a specific sub-domain or perspective, \eg, Deep Research which emphasizes professional report generation from extensive information seeking~\citep{xu2025comprehensive,huang2025deepresearch} or the integration of reasoning and RAG~\citep{liang2025reasoning,gao2025synergizing}, our work comprehensively analyzes the holistic pipeline of search agents, including search structure, optimization, application, evaluation, and challenges. For each part, we provide a thorough analysis of representative works and developing tendencies.

Specifically, this paper is structured as follows: Sec.~\ref {sec:formulation} introduces the task formulation for search agents. \textbf{\textit{How to Search}} in Sec.~\ref{sec:what} presents how agents scale up search turns and utilize complex search structures (\ie, parallel, sequential, and hybrid) to determine query content. \textbf{\textit{How to Optimize}} in Sec.~\ref{sec:how} discusses various optimization methodologies for search agents, including tuning and non-tuning approaches. \textbf{\textit{How to Apply}} in Sec.~\ref{sec:where} delineates the extensive application areas of search agents, encompassing both internal agent enhancements (\eg, reasoning, memory, and tool-use) and external applications (\eg, math, medicine, and finance). \textbf{\textit{How to Evaluate}} in Sec.~\ref{sec:where} introduces evaluation of search agents, covering various datasets and metrics. Finally, Sec.~\ref{sec:challenge} presents current challenges and promising future research directions.

    

\begin{figure*}
    \centering
    \vspace{-10pt}
    \includegraphics[clip,width=\textwidth]{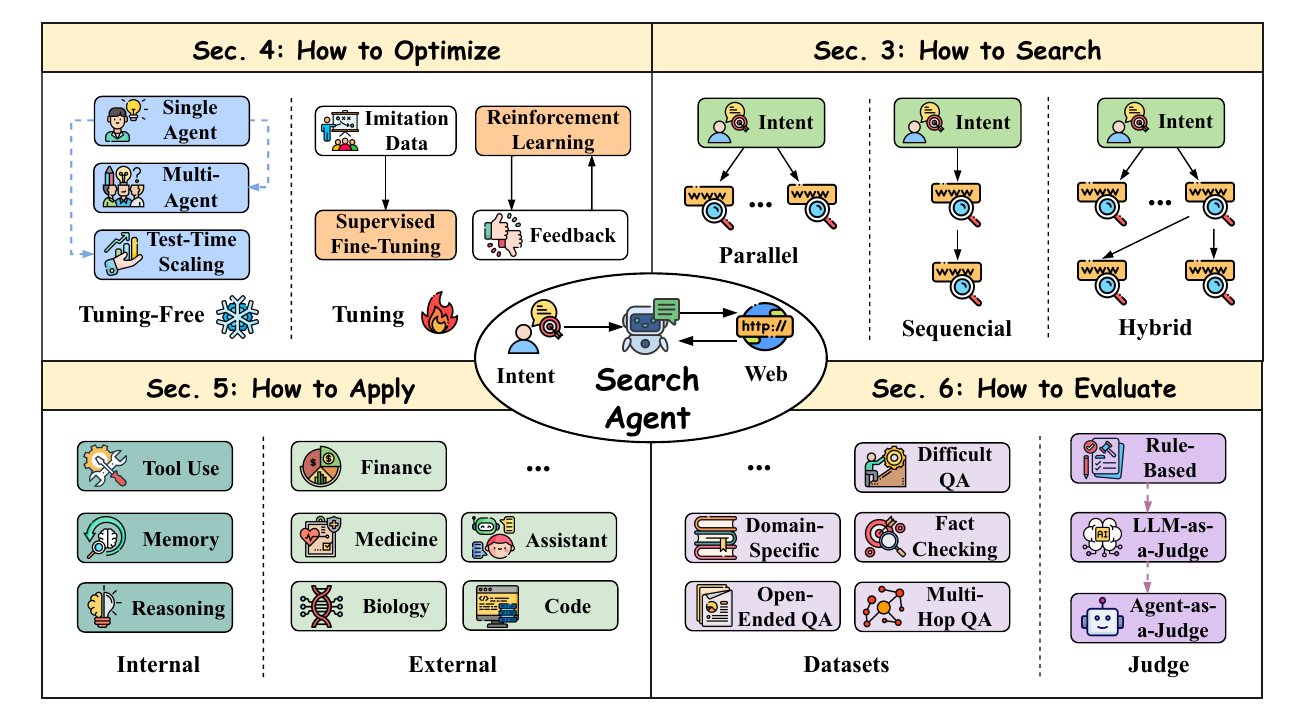}
    \vspace{-20pt}
    \caption{Structural overview of Search Agent -- how to search, how to optimize, how to apply, and how to evaluate.}
    \vspace{-10pt}
    \label{fig:framework}
\end{figure*}

\section{Task Formulation}\label{sec:formulation}
Given a user's intention $q$ and context $C$, a search agent iteratively plans and acts to gather information and fulfill the user's intention.

Upon receiving intention $q$, the agent initiates a planning $\pi_0=\text{Plan}(q, C)$ to conduct an information seeking trajectory. At each step $t$, the agent reflects on its current observation $o_t$ and previous trajectory $t$ and updates its plan $\pi_{t+1}=\text{Reflect}(o_t,h_t,\pi_t)$. It then performs an action $a_{t+1}=\text{Act}(\pi_{t+1})$ (\eg, search for or browse certain content) yielding a new observation $o_t$, \eg, the retrieved information. This process continues until sufficient information is acquired, forming a sequence of observations $O=\{o_1,o_2,\cdots,o_T\}$. From $O$, the agent extracts and ranks the most relevant data into an evidence set $E=\text{Select}(q, O)$ and generates a response $\hat{y}_q=\text{Generate}(q, E)$ to fulfill the user's intention.

\section{How to Search}\label{sec:what}


The core of a search agent lies in its ability to autonomously determine its actions based on user intent and environment context,  deciding when and what to reason or search. This multi-turn process represents a significant shift towards "\textit{scaling up test-time search}." Consequently, the traditional single-query has evolved into dynamic and context-dependent queries, where search queries are decided by sophisticated search structures (parallel, sequential, and hybrid) and search feedback. 


\subsection{Parallel Structure}
Parallel search structures involve reformulating a single query into multiple distinct queries that can be processed simultaneously. This approach is often seen in earlier works, serving as a transitional phase from LLM-enhanced search.

\paragraph{Decomposition-based Parallel Search.}When user intents are complex or vague, direct retrieval often fails. Decomposition-based work tackles this by a planning-execution-verification paradigm, breaking the original intent into smaller sub-queries, which are then executed in parallel and synthesized for a complete answer~\citep{press2022measuring}. This approach focuses primarily on how to decompose the original query better. For instance, \citet{khattab2022demonstrate,shi2024generate,zhang2025credible} leverage strong LLMs through prompting to perform decomposition; \citet{wang2024learning} further learns decomposition strategies from structured knowledge graphs; while \citet{li2023chain,joshi2024reaper} utilize fine-tuned smaller LLMs to generate retrieval plans and queries.

\paragraph{Diversification-based Parallel Search.}Sometimes, a user's intent may correspond to multiple plausible queries. Thus, diversification-based approaches rewrite the original query into a diverse set of queries to be searched in parallel. This strategy ensures that the retrieved content captures a broader range of perspectives and interpretations. For instance, \citet{kostric2024surprisingly} employs beam search to generate multiple candidate queries, while \citet{dhole2024genqrensemble} synthesizes diverse keywords by ensembling various prompts. Other approaches adopt a hybrid strategy, combining standard query rewriting, keyword extraction, and even the use of LLM-generated Pseudo-Answers as queries to enhance retrieval breadth and relevance~\citep{li2024dmqr,abbasiantaeb2024generating,seo2025qa}.


\subsection{Sequential Structure}
Parallel structure determines all search queries in advance, making it less adaptable to unexpected issues that may arise during search. In contrast, the sequential structure is more dynamic and flexible, allowing the agent to decide whether and what to search next based on results and reflections from prior steps. Note that early LLM-enhanced search also involves sequential structures; however, they are typically rule-based retrieval for each sentence or thought, rather than agent-driven mode~\citep{yang2025survey,jiang2023active,shao2023enhancing,trivedi2022interleaving,wang2024rat}.

\paragraph{Reflection-Driven Sequential Search.} This approach commonly employs a loop-based mechanism where the agent performs a search, generates an answer based on search results, and then reflects on its quality and correctness~\citep{hayashi2025iterkey,xiao2025retrieval,zhou2024metacognitive,lee2024planrag,leeagent}. Based on the reflection, the agent initiates further planning and search, iterating until a satisfactory response is achieved.

\paragraph{Proactivity-Driven Sequential Search.}  In this paradigm, the agent decides when to trigger a search and what to search based on the context, \eg, user intent, search results, and previous reasoning. Such dynamic sequential decision-making can be guided by prompts that encode human-designed heuristics, allowing the agent to reason, act, and reflect step by step in response to intermediate outcomes~\citep{li2025search,jiang2025retrieve,huang2025manusearch,wu2025agentic}. Alternatively, it can be learned through fine-tuning, \ie, imitating expert trajectories~\citep{asai2023self,islam2024open,yu2024auto,aksitov2023rest} or exploring the environment autonomously to discover more effective strategies~\citep{jin2025search,song2025r1,chen2025learning,zheng2025deepresearcher,wang2025vrag}.


\subsection{Hybrid Structure}

The hybrid structure combines both parallel and sequential paradigms, enabling exploration along multiple paths simultaneously, increasing the likelihood of covering the correct answer. Based on the underlying structural properties, it can be categorized into tree-based and graph-based structures.

\paragraph{Tree-based Search} Here, each node represents a retrieval step, and at each iteration, multiple successor nodes can be expanded in parallel from a given node. The final answer is then selected or synthesized by aggregating results from various paths to identify the most optimal outcome. Some approaches employ rule-based tree structures, where queries are first decomposed in parallel and then each path is explored independently before merging the results~\citep{zhang2024hierarchical,li2025knowcoder,nguyen2025ma}. Other methods adopt more dynamic strategies by using expansion functions to generate nodes and applying Monte Carlo Tree Search (MCTS) algorithms, where the final answer is selected through mechanisms such as voting~\citep{trinh2025towards,feng2025airrag,tran2024rare} or reward model~\citep{xiong2025mcts,li2024can,ren2025holistically}.

\paragraph{Graph-based Search} Graph-based structures allow arbitrary connections between nodes, enabling the search process to backtrack and revise earlier decisions. Some approaches decompose the problem into a directed acyclic graph (DAG), identifying dependencies between queries and traversing the graph dynamically~\citep{li2024agent}. Others support dynamic node expansion~\citep{chen2024mindsearch,hu2024level} and shrinkage~\citep{teng2025atom}, allowing the agent to adaptively revise its reasoning and search direction.

Recent trends in search structure indicate a shift towards dynamism, from fixed sub-queries to contextually generated ones, and from parallel to sequential and hybrid structures. As shown in Table~\ref{appendix:non-tuning-based} and ~\ref{appendix:tuning-based}, hybrid structures are often preferred in tuning-free settings to cover more search paths and improve performance, while fine-tuned models tend to internalize this flexibility within sequential structures, thereby improving both efficiency and effectiveness.
\section{How to Optimize}\label{sec:how}
 This section explores the key approaches to optimizing search agents, broadly categorized into tuning-free and tuning-based methods.
\subsection{Tuning-Free Approaches}
Tuning-free approaches primarily rely on human knowledge and predefined workflows to guide the agent's actions. 
While basic prompt-driven solutions often involve a single agent, there is a growing trend towards employing multi-agent architectures and test-time scaling to boost performance.
See Table~\ref{appendix:non-tuning-based} for a more detailed comparison.

\paragraph{Single-Agent Architectures.} In this architecture, a single agent handles the entire search process, including planning, query generation, and final answer synthesis. Since it's challenging for a single, prompt-driven agent to control every aspect of a complex process dynamically, these systems often adopt pre-defined, structured workflows. For example, some works involve an \textit{iterative refinement loop}, \ie,  the agent first searches, then generates an answer, evaluates its quality, and finally decides whether to further search. This loop continues until a satisfactory answer is produced~\citep{hayashi2025iterkey,xiao2025retrieval,lee2024planrag,zhou2024metacognitive}. Other works follow the \textit{reflection chain}, where the agent reflects on its current progress or its intermediate results, dynamically adjusting its strategy~\citep{wang2023knowledge,zhang2024hierarchical,xiong2025mcts,li2025knowcoder}.

\paragraph{Multi-Agent Architectures.}While it is challenging for a single agent to handle multiple search tasks, multi-agent architectures decompose complex search tasks and distribute them among specialized agents. Common roles include \textit{planner agent}, control the process, and terminate; \textit{search agent}, gathering evidence from external sources; and \textit{generation agent}, synthesizing the final answer with the collected evidence~\citep{jiang2025retrieve,chen2024mindsearch,huang2025manusearch,hu2024level,ma2025autodata}. Some work also involves \textit{browser agent}~\citep{huang2025manusearch,du2025deepresearch}, \textit{evaluator agent}~\citep{nguyen2025ma,wang2025vidorag,trinh2025robust}, and \textit{memory agent}~\citep{wu2025agentic}.

In terms of process control, most solutions employ a fixed execution order, where one agent passes its output to the next upon completion~\citep{jiang2025retrieve,chen2024mindsearch,huang2025manusearch,hu2024level,wang2025vidorag,trinh2025robust}. Others adopt a centralized supervisor agent that dynamically allocates tasks and invokes other agents~\citep{wu2025agentic,ma2025autodata}. 


\paragraph{Test-Time Scaling.}


Test-time scaling enhances agent performance by allocating more computation during inference to boost task performance, as previously demonstrated by models like OpenAI-o1 and DeepSeek-R1~\citep{zhang2025survey}. Recent work also confirms \textbf{\textit{reasoning-centric scaling}} can significantly improve search agent~\citep{zhang2025web,lee2024planrag,wei2025browsecomp}. 

As search agents interact with external environments, they offer an additional \textbf{\textit{search-centric scaling}}: increasing the number of interactions with external environments for better external knowledge exploration. \citet{xi2025infodeepseek} has observed that the performance of search agents scales smoothly with an increase in the maximum permissible number of search actions. 
Recent research is integrating both reasoning and search scaling, simultaneously deepening reasoning and allowing for more search actions~\citep{feng2025airrag,tran2024rare,jiang2024rag,ren2025holistically}. This combined approach leverages well-known inference scaling techniques such as Self-Consistency (SC), Best-of-N (BoN), and Monte Carlo Tree Search (MCTS) to achieve superior final results.

\subsection{Tuning-based Approaches}
Tuning-based approaches involve training the agent to automatically learn its next actions based on the current context through imitation and exploration. See Table~\ref{appendix:tuning-based} for a more detailed comparison.
\paragraph{Supervised Fine-Tuning (SFT).}
SFT directly trains LLMs on datasets comprising high-quality reasoning and search trajectories or actions, thereby internalizing these capabilities into the model. 
Recent works leverage SFT to enhance either components of search agents, \eg, query rewriting and reflection, or the fully end-to-end reasoning-search pipeline. For \textit{\textbf{component-level tuning}}, annotation is often sourced from expert LLMs~\citep{asai2023self,islam2024open,yu2024auto}. For \textit{\textbf{holistic training}}, synthetic trajectories are generated via LLM-environment interactions and filtered by rejection sampling~\citep{aksitov2023rest,pan2023kwaiagents,zhang2025agent}. 
Typical filtering criteria involve the correctness of the answer judged by an LLM against ground truth~\citep{song2025r1++,li2025webthinker}. More nuanced filtering criteria, including diversity, informativeness, and efficiency, are sometimes employed to further optimize training signal fidelity~\citep{li2025webthinker,zhang2025evolvesearch,sun2025simpledeepsearcher}. 
In some cases, general-purpose SFT data is blended to preserve agents' general capabilities~\cite{lee2025rearag}.


SFT in this context primarily serves a few critical roles: (1) \textit{\textbf{Distillation}}, wherein data generated via elaborate prompting of powerful teacher models is transferred to smaller student models~\citep{jiang2025ras,asai2023self,islam2024open,lee2025rearag,pan2023kwaiagents}; (2) \textit{\textbf{Self-Improvement}}, involving iterative retraining on self-generated, high-quality trajectories filtered by rejection sampling~\citep{wang2025chain,aksitov2023rest,wu2025masksearch}; (3) \textit{\textbf{Preparation for RL training}}, where SFT serves as a crucial warm-up to initialize models with essential operational priors~\citep{zhang2025agent,li2025webthinker,song2025r1++,zhang2025evolvesearch,shi2025pangu}, and also underpins reward model training for downstream reinforcement learning~\citep{luo2025kbqa,xiong2025rag,sun2025rearter}.

\paragraph{Reinforcement Learning (RL).}
RL empowers search agents to learn flexible, optimal behaviors through exploration with environments. While some efforts target optimizing specific components of search agents, \eg, retrieval~\citep{jiang2025deepretrieval,hsu2024grounding} and re-ranking~\citep{xu2025mm}, a growing trend is to integrate RL into end-to-end training of the pipeline, including planning, searching, reflection, and generation~\citep{jin2025search,song2025r1,zheng2025deepresearcher}. Commercial solutions, \eg, OpenAI and Gemini's deep research, have also incorporated proprietary RL implementations. For \textit{\textbf{RL algorithms}}, search agent optimization largely relies on common RL algorithms like PPO~\citep{schulman2017proximal}, GRPO~\citep{shao2024deepseekmath}, and Reinforce++~\citep{hu2025reinforce++}. Though some studies analyze various algorithms for search agents~\citep{song2025r1,jin2025empirical,xiong2025rag,sun2025zerosearch,jin2025search}, there's no current consensus on which is best.


A crucial aspect of optimizing search agents with RL is constructing \textit{\textbf{multi-objective reward functions}}. While format adherence and answer correctness are nearly ubiquitous reward components, other objectives frequently integrated include efficiency~\citep{wang2025vrag,huang2025reinforced}, diversity~\citep{mei20252,dao2025rezero}, evidence quality~\citep{qian2025scent,zhao2025r}, and retrieval gain~\citep{wang2025stepsearch,shi2025search}. Additionally, penalties for redundancy and length are often incorporated to refine behavior~\citep{wang2025stepsearch,wu2025masksearch,song2025r1++}. 
These rewards largely rely on rule-based verification, Outcome-based Reward Models (ORM), and Process-based Reward Models (PRM). Rule-based rewards are applied for verifiable results, \eg, questions with standard answers. ORM typically uses LLMs to judge results without standard answers. PRM assesses utility from each individual step within the search trajectory~\citep{zhang2025process,xiong2025rag,sun2025rearter}.


\paragraph{Mixed Approaches.}
These represent a robust strategy that combines multiple tuning methods. The prevailing methodology involves utilizing SFT as a warm-up phase for the RL stage~\citep{zhang2025agent,li2025webthinker,song2025r1++,zhang2025evolvesearch,shi2025pangu}. This helps the model adapt to the task and correct format, providing a sound initialization for RL training.

Further optimizations implement an iterative training loop between SFT and RL~\citep{zhang2025evolvesearch}, where SFT is optimized with high-quality rollouts from previous RL iterations, providing an enhanced initialization for the next RL training cycle. \citet{shi2025iterative} employs a Generalized Expectation-Maximization framework for iterative trajectory exploration and optimization. \citet{zhang2025agent} integrates contrastive learning to learn when to trigger retrieval more effectively. Furthermore, \citet{wu2025masksearch} introduces a pre-training task, Retrieval-Augmented Mask Prediction, before the SFT and RL stages to boost the model's fundamental capabilities.

\section{How to Apply}\label{sec:where}

Search Agents, with their flexibility and proactivity, have expanded information retrieval far beyond traditional web search. Externally, they conduct in-depth information seeking in diverse domains; internally, they enhance the agent’s capabilities through targeted information search. More applications and comparisons are provided in Table~\ref{appendix:applications}.

\subsection{External Applications}

Search Agents are revolutionizing various industries and applications. A prominent example of this is their seamless integration into chatbots and \textbf{AI assistants}, \eg, OpenAI, Gemini, Perplexity, and Gork. 
A particularly significant development within AI assistants is Deep Research (DR). DR systems are meticulously designed to conduct exhaustive searches across diverse sources, synthesize vast amounts of disparate information, and then present their findings in a well-organized, often professional report. Since 2025, there have been many successful commercial applications, \eg, OpenAI DR~\citep{openai_dr}, Perplexity DR~\citep{Perplexity_dr}, Gemini DR~\citep{gemini_dr}. The field is also thriving with open-source projects like Jina AI node-DeepResearch~\citep{Jina_AI_node-DeepResearch} and Langchain's Open Deep Research~\citep{open_deep_research}, alongside a growing body of academic research work~\citep{yang2025multimodal,singh2025code,li2025webthinker,yuan2025videodeepresearch}.

Beyond AI assistants, Search Agents are also finding fertile ground in a variety of other specialized domains, including \textbf{e-commerce}~\citep{bi2025stepo,lyu2025deepshop,zhao2024let,wang2024macrec}, \textbf{finance}~\citep{li2024agent,lee2024planrag,vaghefi2025ai}, \textbf{code}~\citep{singh2025code,zhang2024codeagent,singh2025code}, \textbf{medicine}~\citep{bhattacharyya2025surgical,chen2025mrd,wang2025medagent}, \textbf{biology}~\citep{liu2024toursynbio,al2025deepseq} and \textbf{chemistry}~\citep{callahan2025agentic,li2025deepsolution}. Besides, Search Agents serve as powerful teaching and \textbf{research assistants} by efficiently collecting materials across various fields~\citep{he2025pasa,brett2025patience,schneider2025collex}. 


\subsection{Internal Applications}
Beyond external sources, a crucial, often overlooked, retrieval source for an agent lies within its internal components: its memory, accumulated experiences, and available tools. Agentic search can be effectively introduced  to significantly enhance its core capabilities:

\paragraph{Tool Use.}As agents gain access to a growing arsenal of tools, identifying the most appropriate one for a given task becomes a pressing challenge. Agentic search offers a solution by multi-turn reasoning and search, enabling more dynamic and precise tool selection~\citep{lumer2024toolshed,du2024anytool,lumer2025scalemcp,xu2024enhancing}.

\paragraph{Memory.}With user interactions accumulating, an agent's memory can become vast and unwieldy. Effectively navigating this information to pinpoint content is another area where agentic search excels. Agentic search can extract queries from complex and ambiguous user intentions and then conduct deep searches within the agent's memory to retrieve highly pertinent information~\citep{xu2025mem,ocker2025grounded,tan2024taskgen}.


\paragraph{Reasoning.}The experiences an agent learns through its self-evolution serve as an invaluable internal search source. Agentic search can dynamically retrieve relevant experiences from this internal reservoir, combining them with externally acquired knowledge to facilitate more robust and insightful reasoning~\citep{wu2025agentic,wang2024rat,wu2025graph,wang2023knowledge}.

Currently, Search Agents are primarily used for Question Answering (QA) tasks, but are rapidly expanding to a broader spectrum of external domains and internal agent capabilities. The ultimate trajectory envisions a seamless and powerful integration of both external knowledge acquisition and internal self-optimization, creating more capable and adaptable agents for complex real-world scenarios.

\section{How to Evaluate}\label{sec:eval}
Evaluating the performance of Search Agents is crucial for understanding their strengths, weaknesses, and overall effectiveness. This section outlines the evaluation process of search agents, including datasets, metrics, and judgment criteria. See Table~\ref{appendix:datasets} for more details.

\subsection{Datasets for Evaluation}
The datasets for evaluating search agents primarily include complex Question-Answering (QA), alongside challenging reasoning problems that require extensive information seeking. 

\paragraph{Closed-ended QA.} Closed-ended questions have specific, definite answers, which are easy to evaluate. However, simple closed-ended QA datasets like NQ~\citep{kwiatkowski2019NQ} and TriviaQA~\citep{joshi2017triviaqa} are often insufficient to effectively evaluate search agents in multi-turn search. Therefore, more challenging closed-ended QA datasets are developed, specifically focusing on multi-hop QA, challenging QA, and fact-checking. \textbf{Multi-hop QA} requires search agents to synthesize information from multiple sources, necessitating iterative information retrieval~\citep{yang2018hotpotqa,ho2020constructing,press2022measuring,trivedi2022musique}. \textbf{Challenging QA} is designed to present genuinely difficult problems, often featuring long-horizon questions~\citep{wei2025browsecomp,zhou2025browsecomp} and involving long-tail knowledge and significant distracting information~\citep{xi2025infodeepseek,pham2025sealqa}. \textbf{Fact-checking task} evaluates a search agent's ability to verify factual claims~\citep{wei2024long,wadden2020fact,jiang2020hover,wang2024mfc}. It inherently involves iterative search, browsing, and comparative analysis of disparate information sources. See more details in Appendix~\ref{sec:closed_ended}.

\paragraph{Open-ended QA Dataset.}
While closed-ended problems with definitive answers, many user inquiries are inherently open-ended, lacking a single, unambiguous correct response. Therefore, it is crucial to assess a search agent's performance on such open-ended questions. Our focus here is primarily on datasets designed for open-ended problems requiring deep information seeking, which is central to the objectives of Deep Research, a subcategory of search agents. These involve delving into broad topics to produce comprehensive, high-quality research reports~\citep{coelho2025deepresearchgym}. Such datasets target multi-perspective or non-factual queries~\citep{rosset2024researchy} and expert-level research task~\citep{du2025deepresearch,bosse2025deep,tan2024proxyqa}, with some incorporating multi-modal queries~\citep{yang2025multimodal}.

\paragraph{Domain-Specific Dataset.}
The evaluation of search agents also involved solving domain-specific tasks with agentic information-seeking abilities. This encompasses two prominent subtypes: (1) Information Seeking in Specialized Fields: This involves iterative retrieval unique to domains like finance~\citep{li2024agent}, business~\citep{chen2025xbench,choubey2025benchmarkingdeepsearchheterogeneous}, Medicine~\citep{pal2022medmcqa,chen2025medbrowsecomp}, and agriculture~\citep{dongre2025mirage}, where search agents must integrate domain-specific knowledge to provide relevant, precise answers. (2) Hard Problems in Specific Domains: These involve highly complex challenges in fields like mathematics~\citep{he2024olympiadbench}, physics~\citep{rein2024gpqa}, and code~\citep{shi2024can}, often resembling expert-level or Olympiad-style problems~\citep{mialon2023gaia,phan2025humanity}. Solving such problems requires not only internal reasoning capabilities but also the ability to leverage external information and prior experiences, making agentic search essential.

\subsection{Metrics \& Judgment}

This section introduces various metrics and judgment approaches for evaluating Search Agents. 

\paragraph{Metrics.}
Most benchmarks assess the performance of search agents based on the \textit{effectiveness} of the final output. For closed-ended QA, fact-checking, and most domain-specific tasks, the dominant metric is typically task success rate, \eg, Exact Match (EM), F1 score, Accuracy, and Pass@k. Sometimes, the evaluation extends beyond the final output to the \textit{intermediate processes}, such as evaluating the quality of the reasoning chain~\citep{wu2024cofca} and retrieval~\citep{su2024bright,yao2023end,wei2024long,eisenschlos2021fool}. Since most benchmarks rely on static search environments, retrieval quality is often measured with standard metrics, \eg, Precision, Recall, NDCG, and MAP, calculated against groundtruth documents. Recent work also explores retrieval quality in dynamic environments~\citep{xi2025infodeepseek}.
.

The evaluation of \textit{open-ended tasks} for Deep Research presents significantly more complexity compared to closed-ended tasks. This complexity arises from the absence of a single "correct" answer, necessitating a nuanced evaluation of the agent's ability to synthesize information comprehensively. Their assessment typically requires multifaceted metrics, \eg, key point coverage~\citep{qi2024long,coelho2025deepresearchgym}, informativeness~\citep{du2025deepresearch}, breadth and depth~\citep{jiang2024into}, coherence, organization~\citep{yang2025multimodal}, readability~\citep{du2025deepresearch}, and citation accuracy~\citep{du2025deepresearch}. Furthermore, some studies adopt Arena-based evaluation (Win Rate)~\citep{chandrahasan2025deep,miroyan2025search}, where outputs from different search agents are presented side by side to human or LLM-based judges, who then determine which output is superior, tied, or inferior. This provides a comparative assessment of performance, particularly useful for nuanced, open-ended tasks where absolute metrics might fall short.

\paragraph{Judge.}
The judge of the above metrics has undergone significant evolution, progressing from simplistic rule-based metrics to more intricate LLM- and Agent-based judging paradigms. Initially, search agents were evaluated using \textbf{Rule-based Judges}, such as metrics Exact Match (EM) and F1 score, against predefined ground truth answers. However, these metrics fail to account for semantic variations, where factually correct answers may be phrased differently. Thus, \textbf{LLM-as-a-Judge} paradigm emerged, leveraging LLMs to assess the accuracy and quality of results. This approach is effective for both closed-ended tasks, where it correlates well with ground truth, and open-ended questions, where evaluation relies on predefined qualitative standards and expert reference reports. As agentic search involves multi-step reasoning and dynamic interaction, the \textbf{Agent-as-a-Judge} paradigm is gaining traction~\citep{gou2025mind2web}. This approach utilizes specialized agents to evaluate the entire search process and its output, providing a deeper assessment than LLM-as-a-Judge. Besides, human evaluation remains the gold standard for nuanced judgment; however, its high cost limits its use, typically focusing on limited samples to validate LLM-based evaluation.

Although the scope of evaluation metrics and the precision of judging methodologies continue to improve, there is a compelling need for more complex and comprehensive evaluation dimensions~\citep{zhu2025evolutionary}. Current benchmarks primarily focus on efficiency and information accuracy, but the core competencies of search agents—particularly their ability to effectively retrieve, synthesize, and discriminate between information—should be central to any evaluation framework. Future evaluation paradigms must expand to include metrics that rigorously assess not only efficiency and accuracy but also source citation reliability and the agent’s capacity to distinguish between credible and unreliable information. This will ensure a more thorough, holistic, and robust assessment of search agents.

\section{Challenges and Future Directions}\label{sec:challenge}

Despite the significant advancements in Search Agent capabilities, several formidable challenges remain, particularly as we push towards more sophisticated and autonomous systems.

\paragraph{Broaden and Fuse Information Sources.} While current search agents primarily leverage browsers and public web data, the next frontier involves integrating more private and proprietary datasets, extending from the agent's external environment to its internal knowledge bases. This integration demands sophisticated methods for combining heterogeneous data formats (text, images, structured data, etc.) and, crucially, resolving conflicts or inconsistencies that arise when information from multiple sources contradicts. Developing robust mechanisms to reconcile conflicting facts or synthesize disparate perspectives is paramount for producing coherent and reliable outputs.

\paragraph{Imperfect Retrieval.}
Search agents frequently operate in environments where retrieved information is imperfect, containing noise, biases, or even outright falsehoods. The internet, for instance, is replete with unreliable information. A significant challenge is to enhance the agent's ability to discern and validate external information, moving beyond mere retrieval to critical evaluation. This necessitates developing advanced information verification techniques and robust fact-checking mechanisms within the agent's pipeline. Improving an agent's "skepticism" and its capacity for critical assessment is vital for elevating the quality and trustworthiness of its generated outputs.

\paragraph{From Text to Multi-Modality.} The majority of current search agents are text-centric. However, the real world is inherently multimodal. A major challenge is to transition search agents from purely text-based understanding to incorporating and processing diverse modalities. This involves two key aspects: first, enhancing the underlying search infrastructure to support better multimodal search queries (e.g., searching for images based on textual descriptions, or videos based on actions). Second, it requires improving the search agent's multimodal comprehension and reasoning abilities -- its capacity to understand, synthesize, and reason across text, images, audio, and video to provide more holistic and contextually rich answers.

\paragraph{Customized Reinforcement Learning.} While general RL algorithms have shown promise, a significant challenge lies in developing customized reinforcement learning algorithms specifically optimized for search agents. The unique characteristics of search tasks, such as long-horizon planning, imperfect feedback, and the knowledge boundary of search agent, often do not align perfectly with standard RL frameworks. This calls for novel algorithms that can effectively manage sparse rewards in iterative search, learn optimal query generation strategies, and make efficient decisions about when and how to interact with external tools or information sources. Tailored RL approaches can lead to more flexible and robust agent behaviors that adapt dynamically to varied search scenarios.

Crucially, reward modeling for search agents presents its own set of complexities. Current RL-based Search Agents often rely on reward signals derived from easily verifiable, closed-ended problems, such as factual QA, where short, definitive answers allow for straightforward validation. However, many real-world user queries are open-ended information-seeking problems, lacking a single, clear-cut answer. Designing effective reward functions and comprehensive reinforcement learning schemes for these nuanced, open-ended scenarios remains a substantial hurdle. This involves defining what constitutes a "good" answer when there's no single ground truth and how to reward the process of sophisticated information discovery and synthesis. Tailored RL approaches and innovative reward structures are essential for robust agent behaviors that adapt dynamically to varied and ambiguous search needs.

\paragraph{Robust Infrastructure.} The ambitious goals for search agents necessitate substantial advancements in underlying infrastructure. A critical challenge is enhancing the efficiency of RL sampling, which can be computationally intensive, to accelerate training. More broadly, optimizing the entire support infrastructure for search agents is crucial. This includes developing high-recall approximate retrieval systems for faster and more relevant information access, implementing priority-aware scheduling to manage complex, concurrent search tasks efficiently, and designing systems that can dynamically schedule requests based on real-time task status, ensuring responsiveness and optimal resource allocation.

\paragraph{Search Agent Self-Evolution.} The ultimate frontier for search agents involves achieving true self-evolution. This presents a profound challenge: enabling agents to continuously learn, adapt, and improve their search strategies and capabilities autonomously over time, without constant human intervention. This involves developing mechanisms for agents to identify their own limitations, generate novel hypotheses about how to improve, and then test those hypotheses through interaction with the environment. Such self-evolving agents would possess an unprecedented capacity for discovery and problem-solving, marking a new era in artificial intelligence.

\section{Conclusion}

The evolution from traditional web search to LLM-enhanced search and Search Agents profoundly transforms information retrieval. These autonomous agents proactively leverage context and diverse sources, transforming search into a proactive, intelligent process. Our survey offers the first systematic analysis, dissecting their mechanisms, optimization, applications, and evaluation. This comprehensive view illuminates the vast potential of search agents for deep information mining and highlights challenges for fully realizing their transformative promise.

\section*{Limitations}
This survey, while comprehensive and systematic, has some limitations. This work primarily focuses on academic research papers, which means it less extensively covers the intricacies of commercial applications. Since companies like OpenAI, Google (Gemini), and Perplexity often do not disclose the specific technical details of their deep research or search agent implementations, there's an inherent gap. This raises a crucial question: are the research directions and observed performance in academic settings truly aligned with the approaches and effectiveness seen in real-world commercial deployments? 

\bibliography{custom}

\appendix
\section{Classification of Closed-ended QA}\label{sec:closed_ended}
\paragraph{Multi-hop QA Dataset.}
Unlike straightforward QA questions, \eg, NQ~\citep{kwiatkowski2019NQ}, TriviaQA~\citep{joshi2017triviaqa}, and MS MARCO~\citep{nguyen2016ms}, Multi-hop QA demands that agents piece together information from multiple sources, necessitating multi-step reasoning and iterative information retrieval. Frequently employed datasets for this purpose include HotpotQA~\citep{yang2018hotpotqa}, 2WikiMultiHopQA~\citep{ho2020constructing}, Bamboogle~\citep{press2022measuring}, and MuSiQue~\citep{trivedi2022musique}. Recent research has further explored implicit multi-hop questions~\citep{geva2021did}, reasoning over information from diverse sources~\citep{tang2024multihop}, and evaluating the reliability of the reasoning chain~\citep{wu2024cofca}. Additionally, some efforts extend multi-hop QA to integrate various reasoning types~\citep{schnitzler2024morehopqa} and multi-modal information~\citep{wang2025vidorag}, pushing the boundaries of complexity.

\paragraph{Challenging QA Dataset.}
With the expansion of LLMs' internal knowledge, many multi-hop questions are now solvable by these models only with their parametric memory, thereby failing to sufficiently engage iterative planning and retrieval capabilities of search agents. To address this, some new challenging benchmarks are emerging, crafted to present genuinely challenging QA problems set within real-world web environments. Examples like BrowseComp~\citep{wei2025browsecomp}, BrowseComp-ZH~\citep{zhou2025browsecomp}, and Mind2Web2~\citep{gou2025mind2web} feature long-horizon problems that require extended search durations. Datasets such as InfoDeepSeek~\citep{xi2025infodeepseek}, ORION~\citep{huang2025manusearch}, and SealQA~\citep{pham2025sealqa} target long-tail knowledge and questions with substantial distracting information. Some method further ensures question difficulty by filtering out the questions that LLMs can easily overcome with a single-turn search~\citep{xi2025infodeepseek}

\paragraph{Fact-Checking Dataset.}
Fact-checking constitutes another critical task for evaluating search agents. The process of verifying factual claims inherently demands iterative search, browsing, and the comparative analysis of disparate information sources. This task is evolving from simple text-based verification to the analysis of long-form factuality~\citep{wei2024long}, unstructured data~\citep{aly2021feverous}, multi-modal content~\citep{wang2024mfc,yao2023end,yang2025realfactbench}, and complex multi-hop fact-checking scenarios~\citep{wadden2020fact,jiang2020hover,eisenschlos2021fool,ostrowski2020multi}, pushing the boundaries of search agent reliability and trustworthiness in sourcing accurate information.

\section{Detailed Challenges and Future Directions}
\label{sec:app_challenge}

Despite the significant advancements in Search Agent capabilities, several formidable challenges remain, particularly as we push towards more sophisticated and autonomous systems.

\paragraph{Broaden and Fuse Information Sources.} While current search agents primarily leverage browsers and public web data, the next frontier involves integrating more private and proprietary datasets, extending from the agent's external environment to its internal knowledge bases. This integration demands sophisticated methods for combining heterogeneous data formats (text, images, structured data, etc.) and, crucially, resolving conflicts or inconsistencies that arise when information from multiple sources contradicts. Developing robust mechanisms to reconcile conflicting facts or synthesize disparate perspectives is paramount for producing coherent and reliable outputs.

\paragraph{Imperfect Retrieval.}
Search agents frequently operate in environments where retrieved information is imperfect, containing noise, biases, or even outright falsehoods. The internet, for instance, is replete with unreliable information. A significant challenge is to enhance the agent's ability to discern and validate external information, moving beyond mere retrieval to critical evaluation. This necessitates developing advanced information verification techniques and robust fact-checking mechanisms within the agent's pipeline. Improving an agent's "skepticism" and its capacity for critical assessment is vital for elevating the quality and trustworthiness of its generated outputs.

\paragraph{From Text to Multi-Modality.} The majority of current search agents are text-centric. However, the real world is inherently multimodal. A major challenge is to transition search agents from purely text-based understanding to incorporating and processing diverse modalities. This involves two key aspects: first, enhancing the underlying search infrastructure to support better multimodal search queries (e.g., searching for images based on textual descriptions, or videos based on actions). Second, it requires improving the search agent's multimodal comprehension and reasoning abilities -- its capacity to understand, synthesize, and reason across text, images, audio, and video to provide more holistic and contextually rich answers.

\paragraph{Customized Reinforcement Learning.} While general RL algorithms have shown promise, a significant challenge lies in developing customized reinforcement learning algorithms specifically optimized for search agents. The unique characteristics of search tasks, such as long-horizon planning, imperfect feedback, and the knowledge boundary of search agent, often do not align perfectly with standard RL frameworks. This calls for novel algorithms that can effectively manage sparse rewards in iterative search, learn optimal query generation strategies, and make efficient decisions about when and how to interact with external tools or information sources. Tailored RL approaches can lead to more flexible and robust agent behaviors that adapt dynamically to varied search scenarios.

Crucially, reward modeling for search agents presents its own set of complexities. Current RL-based Search Agents often rely on reward signals derived from easily verifiable, closed-ended problems, such as factual QA, where short, definitive answers allow for straightforward validation. However, many real-world user queries are open-ended information-seeking problems, lacking a single, clear-cut answer. Designing effective reward functions and comprehensive reinforcement learning schemes for these nuanced, open-ended scenarios remains a substantial hurdle. This involves defining what constitutes a "good" answer when there's no single ground truth and how to reward the process of sophisticated information discovery and synthesis. Tailored RL approaches and innovative reward structures are essential for robust agent behaviors that adapt dynamically to varied and ambiguous search needs.

\paragraph{Robust Infrastructure.} The ambitious goals for search agents necessitate substantial advancements in underlying infrastructure. A critical challenge is enhancing the efficiency of RL sampling, which can be computationally intensive, to accelerate training. More broadly, optimizing the entire support infrastructure for search agents is crucial. This includes developing high-recall approximate retrieval systems for faster and more relevant information access, implementing priority-aware scheduling to manage complex, concurrent search tasks efficiently, and designing systems that can dynamically schedule requests based on real-time task status, ensuring responsiveness and optimal resource allocation.

\paragraph{Search Agent Self-Evolution.} The ultimate frontier for search agents involves achieving true self-evolution. This presents a profound challenge: enabling agents to continuously learn, adapt, and improve their search strategies and capabilities autonomously over time, without constant human intervention. This involves developing mechanisms for agents to identify their own limitations, generate novel hypotheses about how to improve, and then test those hypotheses through interaction with the environment. Such self-evolving agents would possess an unprecedented capacity for discovery and problem-solving, marking a new era in artificial intelligence.

\section{Detailed Tables}
\label{sec:table}
Table~\ref{appendix:non-tuning-based} presents the detailed classification and comparison of non-tuning methods, showing their search structures (parallel, sequential, or hybrid), their detailed sub-structures, whether they adopt multi-agent architectures and test-time scaling (TTS) strategies, as well as their evaluation methods and metrics.

Table~\ref{appendix:tuning-based} summarizes tuning-based methods, comparing their search structures and substructures, evaluation approaches, and tuning strategies. It specifically analyzes reinforcement learning (RL) methods in terms of training algorithms (RL algo.), supervision signals(RL supv.), and reward functions.

Table~\ref{appendix:applications} illustrates the diverse application domains of search agents, including internal capabilities such as memory, reasoning, and tool use, as well as external domains such as mathematics, coding, finance, and healthcare. It covers commercial deployments, open-source projects, and academic research.

Table~\ref{appendix:datasets} compares various datasets, analyzing their categories, scales, modalities, construction methods, evaluation environments, and strategies.

\onecolumn
\centering
\small
\begin{longtable}{>{\RaggedRight\arraybackslash}m{2.3cm}
>{\RaggedRight\arraybackslash}m{1.5cm}
>{\RaggedRight\arraybackslash}m{1.5cm}
>{\RaggedRight\arraybackslash}m{1cm}
>{\RaggedRight\arraybackslash}m{1cm}
>{\RaggedRight\arraybackslash}m{1.5cm}
>{\RaggedRight\arraybackslash}m{3.4cm}
}
\caption{Overview of tuning-free methods. \textit{\textbf{TTS}} is short for \textbf{\textit{Test Time Scaling}}.}
\label{appendix:non-tuning-based}\\
\toprule
\textbf{Model Name} & \textbf{Search Structure} & \textbf{Sub-Structure} & \textbf{Multi-Agent} & \textbf{TTS} & \textbf{Evaluation} & \textbf{Metrics} \\
\midrule
\endfirsthead
\toprule
\textbf{Model Name} & \textbf{Search Structure} & \textbf{Sub-Structure} & \textbf{Multi-Agent} & \textbf{TTS} & \textbf{Evaluation} & \textbf{Metrics} \\
\midrule
\endhead
\endfoot

MAS~\cite{trinh2025robust} & Hybrid & Tree & Yes & No & Rules, LLM & Macro F1 \\
\midrule

AI Search Paradigm~\cite{li2025towards} & Hybrid & Graph & Yes & No & \makecell[l]{Human,\\ Online \\A/B Test} & \makecell[l]{Offline metric:\\ Normalized Win Rate \\ Online metric: \\ Change query rate, \\ Number of page views, \\ Number of daily active users,\\ Dwell time} \\
\midrule

KnowCoder-V2~\cite{li2025knowcoder} & Hybrid & Tree & Yes & No & Rules, LLM & \makecell[l]{QA: Hits@1 \\ Report: \\ Completeness, Thoroughness, \\ Factuality, Coherence, Insight} \\
\midrule

Multimodal DeepResearcher~\cite{yang2025multimodal} & Sequential & Reflection & No & No & Human, LLM & \makecell[l]{Informativeness and Depth, \\ Coherence and Organization, \\ Verifiability, \\ Visualization Quality, \\ Visualization Consistency} \\
\midrule
Agentic Deep Research~\cite{zhang2025web} & Hybrid & Tree & No & Yes & Rules & Accuracy \\
\midrule
AutoData \cite{ma2025autodata} & Hybrid & Graph & Yes & No & Rules & \makecell[l]{ F1 score, \\ Precision, Recall,\\ Task finishing time} \\
\midrule
ManuSearch~\cite{huang2025manusearch} & Hybrid & Tree & Yes & No & LLM & Pass@1 \\
\midrule
Code Researcher~\cite{singh2025code} & Sequential & Proactivity & No & No & Rules & \makecell[l]{Pass,\\ Crash Resolution Rate} \\
\midrule
MA-RAG~\cite{nguyen2025ma} & Hybrid & Tree & Yes & No & Rules & \makecell[l]{QA: Exact Match \\ Fact checking: Accuracy} \\
\midrule
IterKey~\cite{hayashi2025iterkey} & Sequential & Reflection & No & No & Rules & Exact Match, Recall \\
\midrule
ODS~\cite{alzubi2025open} & Sequential & Proactivity & No & No & LLM & Accuracy \\
\midrule
MCTS-RAG~\cite{hu2025mcts} & Hybrid & Tree & No & Yes & Rules & Accuracy \\
\midrule
N/A~\cite{xu2025agentic} & Hybrid & Tree & Yes & No & N/A & N/A \\
\midrule
HG-MCTS~\cite{ren2025holistically} & Hybrid & Tree & No & Yes & Rules & \makecell[l]{String accuracy, ROUGE, \\ Exact Match, F1 score, \\ Cover Exact Match} \\
\midrule
ViDoRAG~\cite{wang2025vidorag} & Sequential & Proactivity & Yes & Yes & Rules, LLM & Accuracy, Recall \\
\midrule
Agentic Reasoning~\cite{wu2025agentic} & Sequential & Proactivity & Yes & Yes & Rules & Accuracy \\
\midrule
FinSearch~\cite{li2024agent} & Hybrid & Graph & No & No & Rules & Accuracy, Processing time \\
\midrule
SolutionRAG~\cite{li2025deepsolution} & Hybrid & Tree & No & Yes & LLM & \makecell[l]{Analytical Score,\\ Technical Score} \\
\midrule
MCTS-KBQA~\cite{xiong2025mcts} & Hybrid & Tree & No & Yes & Rules & \makecell[l]{F1 score, Accuracy, \\ Random Hits@1} \\
\midrule
Search-o1~\cite{li2025search} & Sequential & Proactivity & No & No & -- & Exact Match, F1, Pass@1 \\
\midrule
AirRAG~\cite{feng2025airrag} & Hybrid & Tree & No & Yes & -- & \makecell[l]{ Exact Match,\\ F1 score, Accuracy} \\
\midrule
ReARTeR~\cite{sun2025rearter} & Hybrid & Tree & No & Yes & Rules, LLM & Accuracy \\
\midrule
RetroRAG~\cite{xiao2025retrieval} & Sequential & Loop & No & No & Rules & \makecell[l]{Exact Match,\\ Token-level F1, \\ Precision, Recall} \\
\midrule
Level-Navi Agent~\cite{hu2024level} & Hybrid & Graph & Yes & No & Rules, LLM & \makecell[l]{Correctness Scores, \\ Semantic Similarity Scores, \\ Relevance Scores, \\ Searcher Count} \\
\midrule
RAG-Star~\cite{jiang2024rag} & Hybrid & Tree & No & Yes & Rules & Exact Match, F1 score, Cover Exact Match \\
\midrule
AR-MCTS~\cite{dong2024progressive} & Hybrid & Tree & No & Yes & Rules & Accuracy \\
\midrule
SRSA~\cite{wang2024srsa} & Hybrid & Tree & No & No & LLM & \makecell[l]{Informativeness,\\ Completeness, \\ Novelty, Actionability} \\
\midrule
PlanRAG~\cite{verma2025plan} & Hybrid & Graph & No & No & Rules & Accuracy \\
\midrule
CR-Planner~\cite{li2024can} & Hybrid & Tree & No & Yes & Rules & Accuracy, NDCG@10 \\
\midrule
DRAG, IterRAG~\cite{yue2024inference} & Sequential & Loop & No & Yes & Rules & EM, F1, accuracy \\
\midrule
Agent-G~\cite{leeagent} & Sequential & Proactivity & No & No & Rules & \makecell[l]{Hit, Recall, \\ MRR, Accuracy,\\ Hallucination, Missing} \\
\midrule
Co-STORM~\cite{jiang2024into} & Sequential & Reflection & Yes & No & Human, LLM & \makecell[l]{Report Quality: \\ Relevance, Broad Coverage, \\ Depth, Novelty; \\ Discourse Quality: \\ Novelty, Intent Alignment, \\ No Repetition} \\
\midrule
HiRAG~\cite{zhang2024hierarchical} & Hybrid & Tree & Yes & No & Rules & \makecell[l]{ EM; Token-level F1 score,\\ Precision, Recall} \\
\midrule
MindSearch~\cite{chen2024mindsearch} & Hybrid & Graph & Yes & No & Human, LLM & \makecell[l]{ Depth, Breadth,\\ Factuality, Accuracy} \\
\midrule
ReSP~\cite{jiang2025retrieve} & Sequential & Proactivity & Yes & Yes & Rules & F1 score \\
\midrule
PlanRAG~\cite{lee2024planrag} & Sequential & Loop & No & No & Rules & Accuracy \\
\midrule
STORM~\cite{shao2024assisting} & Sequential & Reflection & No & No & Rules, Human & \makecell[l]{Heading Soft Recall, \\ Heading Entity Recall} \\
\midrule
MetaRAG~\cite{zhou2024metacognitive} & Sequential & Proactivity & Yes & No & Rules & EM, Precision, F1, Recall \\
\midrule
KD-CoT~\cite{wang2023knowledge} & Sequential & Proactivity & Yes & No & Rules & Hits@1, F1 \\
\midrule
SearChain~\cite{xu2024search} & Hybrid & Tree & No & No & Rules & ROUGE, Cover EM \\
\midrule
HiRA~\cite{jin2025decoupled} & Hybrid & Tree & Yes & No & LLM & Accuracy \\

\bottomrule
\end{longtable}
\twocolumn

\onecolumn
\centering
{\fontsize{7pt}{8pt}\selectfont
\begin{longtable}{>{\RaggedRight\arraybackslash}m{2.0cm}
>{\RaggedRight\arraybackslash}m{1.2cm}
>{\RaggedRight\arraybackslash}m{0.7cm}
>{\RaggedRight\arraybackslash}m{0.6cm}
>{\RaggedRight\arraybackslash}m{1.1cm}
>{\RaggedRight\arraybackslash}m{1cm}
>{\RaggedRight\arraybackslash}m{1cm}
>
{\RaggedRight\arraybackslash}m{2cm}
>
{\RaggedRight\arraybackslash}m{2cm}
}
\caption{Overview of tuning-based methods.}
\label{appendix:tuning-based}\\
\toprule
\textbf{Model Name} & \textbf{Search Structure} & \textbf{Sub-structure} & \textbf{Tuning} & \textbf{Training Env.} & \textbf{RL Algo.} & \textbf{RL 
Supv.} & \textbf{RL Reward} &
\textbf{Metrics}\\
\midrule
\endfirsthead
\toprule
\textbf{Model Name} & \textbf{Search Structure} & \textbf{Sub-structure} & \textbf{Tuning} & \textbf{Training Env.} & \textbf{RL Algo.} & \textbf{RL 
Supv.} & \textbf{RL Reward} &
\textbf{Metrics}\\
\midrule
\endhead
\endfoot

HARIS~\cite{hu2025coordinating} & Hybrid & Graph & RL & Simulated & GRPO & Rule+ORM & \makecell[l]{Format\\Answer Accuracy \\Decision Accuracy} & \makecell[l]{F1-score \\ Accuarcy} \\
\midrule
CoRAG~\cite{wang2025chain} & Sequential & Proactive & SFT & Simulated & / & / & \makecell[l]{/} & \makecell[l]{F1-score \\ EM}\\
\midrule
Self-RAG~\cite{asai2023self} & Hybrid & Tree & SFT & Simulated & / & / & / & \makecell[l]{EM, F1, FactScore,\\ MAUVE Citation \\Precision and Recall}\\
\midrule
Open-RAG~\cite{islam2024open} & Hybrid & Tree & SFT & Simulated & / & / & / & \makecell[l]{EM, F1, Recall}\\
\midrule
Auto-RAG~\cite{yu2024auto} & Sequential & Proactive & SFT & Simulated & / & / & / & \makecell[l]{EM, F1, Acc}\\
\midrule
RAS~\cite{jiang2025ras} & Hybrid & Graph & SFT & Simulated & / & / & / & \makecell[l]{EM, F1, ROUGE}\\
\midrule
ReST~\cite{aksitov2023rest} & Sequential & Proactive & mixed & Real-world & ReST & ORM & \makecell[l]{/} & Acc\\
\midrule
Kwai~\cite{pan2023kwaiagents} & Sequential & Proactive & SFT & Real-world & / & / & / & \makecell[l]{ROUGE-L, EM}\\
\midrule
SimpleDeepSearcher \cite{sun2025simpledeepsearcher} & Hybrid & Tree & mixed & Real-world & DPO, Reinforce++ & Rule+ORM & \makecell[l]{Format\\Answer F1} & \makecell[l]{F1, LLM-as-Judge}\\
\midrule
ReaRAG~\cite{lee2025rearag} & Sequential & Proactive & SFT & Real-world & / & / & / & \makecell[l]{F1, EM}\\
\midrule
EXSEARCH~\cite{shi2025iterative} & Sequential & Proactive & RL & Simulated & GEM & PRM & \makecell[l]{Trajectory Quality\\Utility} & \makecell[l]{F1, EM}\\
\midrule
KBQA-o1~\cite{luo2025kbqa} & Hybrid & Tree & SFT & Simulated & / & / & / & \makecell[l]{F1, EM}\\
\midrule
Search-R1~\cite{jin2025search} & Sequential & Proactive & RL & Simulated & PPO, GRPO & Rule+ORM & \makecell[l]{EM} & \makecell[l]{EM}\\
\midrule
DeepNote~\cite{wang2024retriever} & Sequential & Proactive & RL & Simulated & DPO & ORM & \makecell[l]{/} & \makecell[l]{F1, EM, Acc}\\
\midrule
R1-Searcher~\cite{song2025r1} & Sequential & Proactive & RL & Simulated & Reinforce++, GRPO & ORM & \makecell[l]{Format\\Answer Acc} & \makecell[l]{EM, LLM-as-Judge}\\
\midrule
ReSearch~\cite{chen2025learning} & Sequential & Proactive & RL & Simulated & GRPO & ORM & \makecell[l]{Format\\Answer Acc} & \makecell[l]{EM, LLM-as-Judge}\\
\midrule
DeepResearcher \cite{zheng2025deepresearcher} & Sequential & Proactive & RL & Real-world & GRPO & Rule+ORM & \makecell[l]{Format\\Answer F1} & \makecell[l]{F1, LLM-as-Judge}\\
\midrule
AutoCOA~\cite{zhang2025agent} & Sequential & Proactive & Mixed & Simulated & GRPO & Rule+ORM & \makecell[l]{Format\\Answer F1} & \makecell[l]{EM, LLM-as-Judge}\\
\midrule
SWiRL~\cite{goldie2025synthetic} & Sequential & Proactive & RL & Simulated & Policy Gradient & PRM & \makecell[l]{LLM(Gemini)} & \makecell[l]{PM,Acc,F1,EM \\ LLM-as-judge}\\
\midrule
O$^2$-Searcher~\cite{mei20252} & Sequential & Proactive & RL & Simulated & GRPO & ORM & Format, Diversity, Factual & \makecell[l]{EM, F1, LLM\\Finding Similarity}\\
\midrule
ZeroSearch~\cite{sun2025zerosearch} & / & / & Mixed & Simulated & \makecell[l]{PPO,GRPO \\ Reinforce++} & Rule & \makecell[l]{Answer Accuracy} & EM\\
\midrule
StepSearch~\cite{wang2025stepsearch} & Sequential & Proactive & RL & Simulated & PPO & PRM & \makecell[l]{Format,Accuracy \\Search Key\\Information Gains\\Redundancy Penalty} & \makecell[l]{EM, F1}\\
\midrule
VRAG-RL~\cite{wang2025vrag} & Sequential & Proactive & RL & Simulated & GRPO & Rule+ORM & \makecell[l]{Retrieval Efficiency\\Pattern Consistency\\Model-Based outcome} & Acc\\
\midrule
WebThinker~\cite{li2025webthinker} & Sequential & Proactive & RL & Real-world & DPO & PRM & likelihood of Trajectory & LLM-as-Judge\\
\midrule
WebDancer~\cite{wu2025webdancer} & Sequential & Proactive & Mixed & Real-world & DAPO & Rule+ORM & \makecell[l]{Format\\LLM(Answer)} & LLM-as-Judge\\
\midrule
MaskSearch~\cite{wu2025masksearch} & Sequential & Proactive & Mixed & Simulated & DAPO & ORM & \makecell[l]{LLM(Answer)} & Token-level Recall\\
\midrule
DeepDiver~\cite{shi2025pangu} & Sequential & Proactive & Mixed & Real-world & GRPO & Rule+ORM & \makecell[l]{Format, Accuracy,\\Extra Search \\Call Rewards,\\Loose\&Strict Scheduler} & LLM-as-Judge\\
\midrule
R1-Searcher++~\cite{song2025r1++} & Sequential & Proactive & Mixed & Simulated & Reinforce++ & ORM & \makecell[l]{Format\\Answer\\Group (Less \\Retrieval)} & \makecell[l]{F1, LLM-as-Judge}\\
\midrule
ReasonRAG~\cite{zhang2025process} & Sequential & Proactive & RL & Simulated & DPO & PRM & \makecell[l]{Shortest Path \\ Reward Estimation} & \makecell[l]{EM, F1}\\
\midrule
ConvSearch-R1~\cite{zhu2025convsearch} & Sequential & Proactive & Mixed & Simulated & GRPO & ORM & \makecell[l]{Rank-Incentive\\ Reward} & MRR,NDCG,Recall\\
\midrule
EvolveSearch~\cite{zhang2025evolvesearch} & Sequential & Proactive & Mixed & Real-world & GRPO & ORM & \makecell[l]{Format\\LLM(Answer)} & LLM-as-Judge\\
\midrule
Inforage~\cite{qian2025scent} & Sequential & Proactive & Mixed & Simulated & PPO & PRM & \makecell[l]{Outcome\\Information Gain\\Efficiency} & EM\\
\midrule
AutoRefine~\cite{shi2025search} & Sequential & Proactive & RL & Simulated & GRPO & ORM+PRM & \makecell[l]{Answer F1\\Retrieval EM} & EM\\
\midrule
MMOA-RAG~\cite{chen2025improving} & Hybrid & Tree & Mixed & Simulated & MAPPO & ORM & \makecell[l]{Answer F1\\various penalty} & EM,F1,Acc\\
\midrule
R-search~\cite{zhao2025r} & Sequential & Proactive & RL & Simulated & GRPO & ORM & \makecell[l]{Answer F1,\\Model-based\\ Evidence,\\EM,Format} & EM,F1\\
\midrule
ReZero~\cite{dao2025rezero} & Sequential & Proactive & RL & Simulated & GRPO & \makecell[l]{Rule+ORM\\ PRM} & \makecell[l]{Acc,Format,Search \\ (Retry/Acc/Diversity)} & Acc\\
\midrule
s3~\cite{jiang2025s3} & Sequential & Proactive & RL & Simulated & PPO & ORM & \makecell[l]{Gain Beyond RAG} & Span Match, LLM-as-Judge\\
\midrule
/~\cite{jin2025empirical} & Sequential & Proactive & RL & Simulated & PPO/GRPO & ORM & \makecell[l]{Format\\Answer F1} & F1\\
\midrule
Re²Search++~\cite{xiong2025rag} & Sequential & Proactive & RL & Simulated & DPO/PPO & PRM & \makecell[l]{Simulated} & EM,F1,Acc, LLM-as-Judge\\
\midrule
ReARTeR~\cite{sun2025rearter} & Sequential & Proactive & RL & Simulated & / & PRM & \makecell[l]{Monte Carlo Score} & Acc, LLM-as-Judge\\
\midrule
SmartRAG~\cite{gao2024smartrag} & Sequential & Proactive & Mixed & Real-world & PPO & ORM & \makecell[l]{Answer Reward \\ Retrieval Penalty} & EM, F1, Hit\\
\midrule
$\mathnormal{\beta}$-GRPO~\cite{wu2025search} & Sequential & Proactive & RL & Simulated & $\mathnormal{\beta}$-GRPO
 & PRM & \makecell[l]{Min Prob of \\ Search tags} & EM\\
\midrule
DeepRAG~\cite{guan2025deeprag} & Hybrid & Binary Tree & Mixed & Simulated & GRPO & ORM & \makecell[l]{Answer acc\\Retrieval Cost} & EM, F1\\
\midrule
IEKA~\cite{huang2025reinforced} & Sequential & Proactive & RL & Simulated & GRPO & ORM & \makecell[l]{Answer EM\\Retrieval Counts\\ Penalty} & EM, Search Valid Rate\\
\midrule
KunLunBaizeRAG \cite{li2025kunlunbaizerag} & Sequential & Proactive & Mixed & Simulated & DAPO & ORM & \makecell[l]{Answer EM\\Format,Length\\Search Effiency} & EM, LLM-as-Judge\\
\midrule
RAG-R1~\cite{tan2025ragr1incentivizesearch} & Parallel & Chain & Mixed & Simulated & PPO & ORM & Answer EM & EM \\
\midrule
Web-Sailor~\cite{li2025websailornavigatingsuperhumanreasoning} & Hybrid & Graph & Mixed & Real-world & DUPO & ORM & \makecell[l]{Format \\ Answer F1} & \makecell[l]{Pass@1 \\ LLM-Acc} \\

\bottomrule
\end{longtable}
}
\twocolumn

\onecolumn
\centering
\small
\setlength{\tabcolsep}{7pt}
\begin{longtable}{>{\centering\arraybackslash}m{4cm}
>{\centering\arraybackslash}m{1cm}
>{\centering\arraybackslash}m{3cm}
>{\centering\arraybackslash}m{1.7cm}
>{\centering\arraybackslash}m{1cm}
>{\centering\arraybackslash}m{2cm}}
\caption{Different applications of Search Agents.}
\label{appendix:applications}\\
\toprule
\textbf{Model Name} & \textbf{Scope} & \textbf{Domain} & \textbf{Use Cases} & \textbf{Tuning} & \textbf{Multi-Agent}\\
\midrule
\endfirsthead
\toprule
\textbf{Model Name} & \textbf{Scope} & \textbf{Domain} & \textbf{Use Cases} & \textbf{Tuning} & \textbf{Multi-Agent}\\
\midrule
\endhead
\endfoot
    
OpenAI Deep Research~\cite{openai_dr} & external & AI assistant & Commercial  & Yes & Yes \\
\midrule
\begin{tabular}[c]{c}Perplexity Deep Research\\~\cite{Perplexity_dr}\end{tabular} & external & AI assistant & Commercial  & Yes & Unknown \\
\midrule
Gemini Deep Research~\cite{gemini_dr} & external & AI assistant & Commercial  & Yes & Unknown \\
\midrule
Grok DeepSearch~\cite{gork_dr} & external & AI assistant & Commercial  & Yes & Unknown \\
\midrule
\begin{tabular}[c]{c}Researcher agent in Copilot\\~\cite{Researcher_agent_in_Microsoft_365_Copilot}\end{tabular} & external & AI assistant & Commercial  & Yes & Unknown \\
\midrule
Kimi Researcher~\cite{Kimi_Researcher} & external & AI assistant & Commercial  & Yes & No \\
\midrule
Elicit~\cite{Elicit_dr} & external & AI assistant & Commercial  & Yes & No \\
\midrule
Consensus~\cite{Consensus_dr} & external & AI assistant & Commercial  & Yes & No \\
\midrule
Manus~\cite{Manus_dr} & external & AI assistant & Commercial  & Yes & Yes \\
\midrule
node-DeepResearch~\cite{Jina_AI_node-DeepResearch} & external & AI assistant & Open Source Project  & Yes & Unknown \\
\midrule
open deep research~\cite{open_deep_research} & external & AI assistant & Open Source Project  & Yes & Yes \\
\midrule
GPT Researcherh~\cite{GPT_Researcher} & external & AI assistant & Open Source Project  & Yes & Yes \\
\midrule
\begin{tabular}[c]{c}Open-source DeepResearch\\~\cite{Open-sourc_DeepResearch}\end{tabular} & external & AI assistant & Open Source Project  & Yes & Yes \\
\midrule
Open Deep Research~\cite{Open_Deep_Research_dzhng} & external & AI assistant & Open Source Project  & Yes & Unknown \\
\midrule
\begin{tabular}[c]{c}open-deep-research\\~\cite{open-deep-research}\end{tabular} & external & AI assistant & Open Source Project  & Yes & No \\
\midrule
AgenticSeek~\cite{AgenticSeek} & external & AI assistant & Open Source Project  & Yes & Yes \\
\midrule
DeerFlow~\cite{DeerFlow} & external & AI assistant & Open Source Project  & Yes & Yes \\
\midrule
OpenManus~\cite{OpenManus} & external & AI assistant & Open Source Project  & Yes & Yes \\
\midrule
SimpleDoc~\cite{jain2025simpledoc} & external & Document Visual & Research  & Yes & Yes \\
\midrule
StePO-Rec~\cite{bi2025stepo} & external & E-commerce & Research  & No & No \\
\midrule
DeepShop~\cite{lyu2025deepshop} & external & E-commerce & Research  & Yes & No \\
\midrule
ARAG~\cite{maragheh2025aragagenticretrievalaugmented} & external & Recommendation & Research  & No & Yes \\
\midrule
MACRec~\cite{wang2024macrec} & external & Recommendation & Research  & No & Yes \\
\midrule
ToolRec~\cite{zhao2024let} & external & Recommendation & Research  & No & Unknown \\
\midrule
PUMA~\cite{cai2025large} & external & E-commerce & Research  & No & Unknown \\
\midrule
FinSearch~\cite{li2024agent} & external & Finance & Research  & Yes & Yes \\
\midrule
PlanRAGh~\cite{lee2024planrag} & external & Business & Research  & Yes & Unknown \\
\midrule
Glass-Box Agent~\cite{vaghefi2025ai} & external & Finance & Research  & Yes & Yes \\
\midrule
Code Researcher~\cite{singh2025code} & external & Code & Research  & Yes & No \\
\midrule
ARCeR~\cite{lupinacci2025arcer} & external & Code & Research  & Yes & No \\
\midrule
CodeAgent~\cite{zhang2024codeagent} & external & Code & Research  & Yes & No \\
\midrule
ARCS~\cite{bhattarai2025arcs} & external & Code & Research  & Yes & No \\
\midrule
\begin{tabular}[c]{c}Agentic Retrieval-Augmented\\~\cite{ravuru2024agentic}\end{tabular} & external & Time Series Analysis & Research  & No & Yes \\
\midrule
Agentic RAG~\cite{spielberger2025retrieval} & external & Topic Modeling & Research  & No & No \\
\midrule
DeRetSyn~\cite{bhattacharyya2025surgical} & external & Medicine & Research  & No & No \\
\midrule
MRD-RAG~\cite{chen2025mrd} & external & Medical Diagnosis & Research  & No & No \\
\midrule
MedAgent-Pro~\cite{wang2025medagent} & external & Medical Diagnosis & Research  & No & No \\
\midrule
\begin{tabular}[c]{c}CBMs and Multi-Agent RAG\\~\cite{tusfiqur2024towards}\end{tabular} & external & Medicine & Research  & No & Yes \\
\midrule
DeepSeq~\cite{al2025deepseq} & external & Biology & Research  & Yes & No \\
\midrule
TourSynbio-Search~\cite{liu2024toursynbio} & external & Biology & Research  & Yes & Yes \\
\midrule
\begin{tabular}[c]{c}LLM-based system\\~\cite{brett2025patience}\end{tabular} & external & Scientific Reserach & Research  & Yes & Yes \\
\midrule
PaSa~\cite{he2025pasa} & external & Scientific Reserach & Research  & Yes & Yes \\
\midrule
CollEX~\cite{schneider2025collex} & external & Scientific Reserach & Research  & Yes & Yes \\
\midrule
\begin{tabular}[c]{c}Claude 3.7 Sonnet\\~\cite{berkane2025llm}\end{tabular} & external & Scientific Reserach & Research  & Yes & No \\
\midrule
IoT-ASE~\cite{elewah2025agentic} & external & Internet of Things & Research  & Yes & Yes \\
\midrule
CRAG-MoW~\cite{callahan2025agentic} & external & Chemistry & Research  & No & Yes \\
\midrule
\begin{tabular}[c]{c}AAWN and Graph\\RAG Framework\\~\cite{srinivas2024accelerating}\end{tabular} & external & Chemistry & Research  & No & Yes \\
\midrule
SolutionRAG~\cite{li2025deepsolution} & external & Engineering & Research  & No & No \\
\midrule
\begin{tabular}[c]{c}Agentic Multimodal RAG\\~\cite{elahi2025one}\end{tabular} & external & Water Resources & Research  & No & Yes \\
\midrule
CMI~\cite{thakrar2025cultivating} & external & Medical Diagnosis & Research  & No & No \\
\midrule
AIDE~\cite{toledo2025airesearchagen} & external & Code & Research  & No & No \\
\midrule
Toolshed~\cite{lumer2024toolshed} & internal & Tool Use & Research  & No & No \\
\midrule
AnyTool~\cite{du2024anytool} & internal & Tool Use & Research  & No & Yes \\
\midrule
ScaleMCP~\cite{lumer2025scalemcp} & internal & Tool Use & Research  & No & No \\
\midrule
TR-Feedback~\cite{xu2024enhancing} & internal & Tool Learning & Research  & No & No \\
\midrule
TaskGen~\cite{tan2024taskgen} & internal & Memory & Research  & No & Yes \\
\midrule
A-MEM~\cite{xu2025mem} & internal & Memory & Research  & No & No \\
\midrule
\begin{tabular}[c]{c}Grounded Memory System\\~\cite{ocker2025grounded}\end{tabular} & internal & Memory & Research  & Yes & No \\
\midrule
Agentic Reasoning~\cite{wu2025agentic} & internal & Memory & Research  & Yes & Yes \\
\midrule
AGENT KB~\cite{tang2025agentkb} & internal & Memory & Research  & No & Yes \\
\midrule
RAT~\cite{wang2024rat} & internal & Reasoning & Research  & Yes & No \\
\midrule
KG-RAR~\cite{wu2025graph} & internal & Reasoning & Research  & Yes & No \\
\midrule
KD-CoT~\cite{wang2023knowledge} & internal & Reasoning & Research  & No & No \\
\midrule
Search-o1~\cite{li2025search} & internal & Reasoning & Research  & Yes & No \\
\midrule
CR-Planner~\cite{li2024can} & internal & Reasoning & Research  & Yes & No \\
\midrule
AR-MCTS~\cite{dong2024progressive} & internal & Reasoning & Research  & Yes & No \\
\midrule
SWiRL~\cite{goldie2025synthetic} & internal & Reasoning & Research  & Yes & No \\
\midrule
WebThinker~\cite{li2025webthinker} & internal & Reasoning & Research  & Yes & No \\
\bottomrule

\end{longtable}
\normalsize
\twocolumn

\onecolumn
\centering
\setlength{\LTleft}{0pt}
\setlength{\LTright}{0pt}
\small
\setlength{\tabcolsep}{4pt}
\begin{longtable}{
>{\centering\arraybackslash}m{3cm}
>{\centering\arraybackslash}m{1.6cm}
>{\centering\arraybackslash}m{1.3cm}
>{\centering\arraybackslash}m{1.3cm}
>{\centering\arraybackslash}m{1.3cm}
>{\centering\arraybackslash}m{1.5cm}
>{\centering\arraybackslash}m{1.5cm}
>{\centering\arraybackslash}m{2.2cm}}
\caption{Overview of datasets.}
\label{appendix:datasets}\\
\toprule
\textbf{Dataset Name} & \textbf{Category} & \textbf{Scale} & \textbf{Modality} & \textbf{Constr-uction} & \textbf{Environ-ments} & \textbf{Evaluation} & \textbf{Metrics}\\
\midrule
\endfirsthead
\toprule
\textbf{Dataset Name} & \textbf{Category} & \textbf{Scale} & \textbf{Modality} & \textbf{Constr-uction} & \textbf{Environ-ments} & \textbf{Evaluation} & \textbf{Metrics}\\
\midrule
\endhead
\endfoot

\begin{tabular}[c] {c}HotpotQA\\~\cite{yang2018hotpotqa}\end{tabular} & Multi-hop QA & 113k & Text  & Rules, Manual & Static & Rules & EM, F1 \\
\midrule
\begin{tabular}[c]{c} 2WikiMultiHopQA\\~\cite{ho2020constructing}\end{tabular} & Multi-hop QA & 192606 & Text  & Rules, Manual & Static & Rules, Human, LLM & EM,F1 \\
\midrule
\begin{tabular}[c]{c} Bamboogle\\~\cite{press-etal-2023-measuring}\end{tabular} & Multi-hop QA & 8600 & Text  & Rules, Manual & Static & Human, LLM & Acc, Gap Ratio, Compositionality \\
\midrule
\begin{tabular}[c]{c} MuSiQue-Ans\\~\cite{trivedi2022musique}\end{tabular} & Multi-hop QA & 25000 & Text  & Manual & N/A & Human & EM,F1 \\
\midrule
\begin{tabular}[c]{c} StrategyQA\\~\cite{geva2021strategyqa}\end{tabular} & Multi-hop QA & 2780 & Text  & Manual & N/A & Human & Acc \\
\midrule
\begin{tabular}[c]{c} FRAMES \\~\cite{krishna2024frames}\end{tabular} & Multi-hop QA & 824 & Text  & Manual & Dynamic & Human & Acc \\
\midrule
\begin{tabular}[c]{c} MultiHop-RAG\\~\cite{tang2024multihoprag}\end{tabular} & Multi-hop QA & / & Text  & Manual & Dynamic & Human & N/A \\
\midrule
\begin{tabular}[c]{c} HoVer\\~\cite{jiang2020hover}\end{tabular} & Fast-Checking & / & Text  & Manual & Dynamic & Human & Acc \\
\midrule
\begin{tabular}[c]{c} FanOutQA\\~\cite{zhu2024fanoutqa}\end{tabular} & Multi-hop QA & 8339 & Text  & Manual & Dynamic & Human & Acc \\
\midrule
\begin{tabular}[c]{c} Web24\\~\cite{hu2025levelnavi}\end{tabular} & Others & / & Text  & Manual & Dynamic & Human & N/A \\
\midrule
\begin{tabular}[c]{c} ViDoSeek\\~\cite{zhang2025vidorag}\end{tabular} & Others & / & Text, Image  & Manual & Dynamic & Human & N/A \\
\midrule
\begin{tabular}[c]{c} MoreHopQA\\~\cite{liu2024morehopqa}\end{tabular} & Multi-hop QA & 1118 & Text  & Manual & N/A & Human & Acc, Reasoning Step Acc \\
\midrule
\begin{tabular}[c]{c} CofCA\\~\cite{wu2025cofca}\end{tabular} & Multi-hop QA & 1800 & Text  & Manual & Static & Human & EM, F1, PM, Reasoning Chain Acc\\
\midrule
\begin{tabular}[c]{c}BrowseComp\\~\cite{wei2025browsecomp}\end{tabular} & Challeging QA & 1266 & Text  & Manual & Dynamic & LLM & Acc, CE \\
\midrule
\begin{tabular}[c]{c}InfoDeepSeek\\~\cite{xi2025infodeepseek}\end{tabular} & Challeging QA & 245 & Text  & Manual & Dynamic & LLM, Human & AnsAcc, InfoAcc, EffEvidUtil, InfoCompactness \\
\midrule
\begin{tabular}[c]{c}ORION\\\cite{huang2025manusearch}\end{tabular} & Challeging QA & 310 & Text  & LLM, Manual & Dynamic & LLM & Pass@1 Acc \\
\midrule
\begin{tabular}[c]{c}BrowseComp-ZH\\\cite{zhou2025browsecomp}\end{tabular} & Challeging QA & 289 & Text  & Manual & Dynamic & LLM, Human & Acc, CE \\
\midrule
\begin{tabular}[c]{c}PopQA\\\cite{mallen2022not}\end{tabular} & Challeging QA & 14k & Text  & Rules & Static & Rules & Acc \\
\midrule
\begin{tabular}[c]{c} WebPuzzle \\ \cite{shi2025pangu} \end{tabular}
 & Challeging QA & 24275 & Text  & LLM, Manual & Dynamic & LLM & Acc, CE \\
\midrule
\begin{tabular}[c]{c} BLUR \\ \cite{ch2025browsing} \end{tabular}
 & Challeging QA & 573 & Text, Image, Video, Audio  & Manual & Dynamic & Rules, LLM & Acc \\
\midrule
\begin{tabular}[c]{c} BRIGHT \\ \cite{su2024bright} \end{tabular}
 & Challeging QA & 1384 & Text & LLM, Manual & Dynamic & Rules & nDCG@10, Precision@10, Recall@10 \\
\midrule
\begin{tabular}[c]{c} SealQA \\ \cite{pham2025sealqa} \end{tabular}
 & Challeging QA & 619 & Text & LLM, Manual & Dynamic & LLM & Acc \\
\midrule
\begin{tabular}[c]{c} MMSearch \\ \cite{jiang2024mmsearch} \end{tabular}
 & Challeging QA & 300 & Text, Image & Manual & Dynamic & Rules & F1, ROUGE-L, BLEU-1, Acc \\
\midrule
\begin{tabular}[c]{c} ScholarSearch \\ \cite{zhou2025scholarsearchbenchmarkingscholarsearching} \end{tabular}
 & Challeging QA & 223 & Text & LLM, Manual & Dynamic & LLM & Acc \\
\midrule
\begin{tabular}[c]{c} Mind2Web \\ 2\cite{gou2025mind2web} \end{tabular}
 & Challeging QA & 130 & Text, Image & LLM, Manual & Dynamic & LLM, Human & PartComp, SuccRate, Pass@3, AvgTime, AnsLen \\
\midrule
\begin{tabular}[c]{c} O2-QA \\ \cite{mei20252} \end{tabular}
 & Open-ended QA & 300 & Text & LLM, Manual & Static & Rules, LLM & F1, EM, LFS \\
\midrule
\begin{tabular}[c]{c} Researchy Questions \\ \cite{rosset2024researchy} \end{tabular}
 & Open-ended QA & ~96k & Text & \begin{tabular}[c]{c}Rules,\\LLM,\\Manual\end{tabular} & Static & LLM & Acc, Score \\
\midrule
\begin{tabular}[c]{c} MultimodalReportBench \\ \cite{yang2025multimodal} \end{tabular} & Open-ended QA & 100 & Text, Image & Manual & Dynamic & LLM, Human & \begin{tabular}[c]{c}InfoDepth,\\OrgCoh, \\Verif, VisQual,\\ VisCons\end{tabular} \\
\midrule
\begin{tabular}[c]{c} DeepResearchGym \\ ~\cite{coelho2025deepresearchgym} \end{tabular}
 & Open-ended QA & unknown & Text & Rules & Static & LLM & KP-Rec, KP-Contra, Cit-Prec, Cit-Rec, Clarity, Insight \\
\midrule
\begin{tabular}[c]{c} Deep Research Bench \\ ~\cite{bosse2025deep} \end{tabular}
 & Open-ended QA & 89 & Text & Manual & Static & \begin{tabular}[c]{c}Rules,\\LLM,\\Human\end{tabular} & \begin{tabular}[c]{c}BinScore, \\Recall(Rec),\\ F1, AbsDiff\end{tabular} \\
\midrule
\begin{tabular}[c]{c} DeepResearch Bench \\ ~\cite{du2025deepresearch} \end{tabular}
 & Open-ended QA & 100 & Text & Manual & Static & Rules, LLM & \begin{tabular}[c]{c}Comprehensiveness, \\Insight/Depth, \\InstFollowing ...\end{tabular} \\
\midrule
\begin{tabular}[c]{c} WildSeek \\ ~\cite{jiang2024into} \end{tabular}
 & Open-ended QA & 100 & Text & \begin{tabular}[c]{c}Rules,\\LLM,\\ Manual\end{tabular} & Dynamic & \begin{tabular}[c]{c}Rules,\\LLM,\\ Human\end{tabular} & Relevance, Breadth, Depth, Novelty, Consistency... \\
\midrule
\begin{tabular}[c]{c} ProxyQA \\ ~\cite{tan2024proxyqa} \end{tabular}
 & Open-ended QA & 100 & Text & Manual & N/A & LLM & Acc, SelfAR, HumAR \\
\midrule
\begin{tabular}[c]{c} Long2RAG \\ ~\cite{qi2024long} \end{tabular}
 & Open-ended QA & 280 & Text & LLM, Manual & Dynamic & LLM & KP-Rec, KP-F1,  KP-Precision(KP-Pre) \\
\midrule
\begin{tabular}[c]{c} Mocheg \\  \cite{yao2023end}\end{tabular}
 & Fact-Checking & 61593 & Text, Image & Rules, Manual & Static & Rules & Prec, Rec, NDCG, MAP, S-Rec \\
\midrule
\begin{tabular}[c]{c} MFC-Bench \\ ~\cite{wang2024mfc} \end{tabular}
 & Fact-Checking & 35k & Text, Image & \begin{tabular}[c]{c}Rules,\\LLM,\\Manual\end{tabular} & Static & Rules & Acc, F1 \\
\midrule
\begin{tabular}[c]{c} RealFactBench \\ ~\cite{yang2025realfactbench} \end{tabular}
 & Fact-Checking & 6k & Text, Image & LLM, Manual & Dynamic & Rules & F1, MCC, UnkRate, ExpQuality \\
\midrule
\begin{tabular}[c]{c} LongFact \\ ~\cite{wei2024long} \end{tabular}
 & Fact-Checking & 2280 & Text & LLM, Manual & Dynamic & LLM & F1@K, Prec, Rec \\
\midrule
\begin{tabular}[c]{c} PolitiHop \\ ~\cite{ostrowski2020multi} \end{tabular}
 & Fact-Checking & 500 & Text & Manual & Static & Rules & macro-F1, Acc, F1, Prec, FEVER score \\
\midrule
\begin{tabular}[c]{c} FM2 \\ ~\cite{eisenschlos2021fool} \end{tabular}
 & Fact-Checking & 12968 & Text & Manual & Static & Rules & R-Prec, Rec@5, Rec@10, Acc, F1, FEVER score \\
\midrule
\begin{tabular}[c]{c} HoVer \\ ~\cite{jiang2020hover} \end{tabular}
 & Fact-Checking & 26171 & Text & Manual & Static & Rules & Acc, F1, EM, R-Prec, Rec@k, HOVER Score \\
\midrule
\begin{tabular}[c]{c} SCIFACT \\ \textbf{} \end{tabular}
 & Fact-Checking & 1409 & Text & Rules, Manual & Static & Rules & \begin{tabular}[c]{c}F1, Acc, Rec, \\FEVER Score\end{tabular} \\
\midrule
\begin{tabular}[c]{c} EX-FEVER \\ ~\cite{ma2023ex} \end{tabular}
 & Fact-Checking & 60k+ & Text & Manual & Static & Rules & EM, Hit@k, Acc, Rouge score \\
\midrule
\begin{tabular}[c]{c} FEVEROUS \\ ~\cite{aly2021feverous} \end{tabular}
 & Fact-Checking & 87026 & Text, Table & Manual & Static & Rules & FEVER Score, EM, Hit@k, Acc, F1, Rec \\
\midrule
\begin{tabular}[c]{c} FactBench \\ ~\cite{bayat2024factbench} \end{tabular}
 & Fact-Checking & 1000 & Text & \begin{tabular}[c]{c}Rules,\\LLM,\\Manual\end{tabular} & Dynamic & LLM & Hallucination Score, Fact-Prec, Acc, CHumJ \\
\midrule
\begin{tabular}[c]{c} FinSearchBench-24 \\ ~\cite{li2024agent} \end{tabular}
 & Domain-Specific & 1500 & Text & \begin{tabular}[c]{c}Rules,\\LLM,\\Manual\end{tabular} & Dynamic & Rules & Acc, Time \\
\midrule
\begin{tabular}[c]{c} MIRAGE \\ ~\cite{dongre2025mirage} \end{tabular}
 & Domain-Specific & 35306 & Text, Image & LLM, Manual & Static & LLM & Reasoning Score, Acc, IdentAcc, Relevance, Completeness... \\
\midrule
\begin{tabular}[c]{c} xbench \\ ~\cite{chen2025xbench} \end{tabular}
 & Domain-Specific & 100 & Text & Manual & Dynamic & LLM & Matching Rate, AccScore\\
\midrule
\begin{tabular}[c]{c} SolutionBench \\ ~\cite{li2025deepsolution} \end{tabular}
 & Domain-Specific & 950 & Text & LLM, Manual & Static & LLM & Analytical Score, Technical Score \\
\midrule
\begin{tabular}[c]{c} DQA \\ ~\cite{lee2024planrag} \end{tabular}
 & Domain-Specific & 301 & Text & Rules & Static & Rules & Acc \\
\midrule
\begin{tabular}[c]{c} MedMCQA \\ ~\cite{pal2022medmcqa} \end{tabular}
 & Domain-Specific & 193k+ & Text & Rules, Manual & Static & LLM & Acc \\
\midrule
\begin{tabular}[c]{c} MedBrowseCom \\ ~\cite{chen2025medbrowsecomp} \end{tabular}
 & Domain-Specific & 1k+ & Text & Manual & Dynamic & LLM, Human & Acc, ECErr \\
\midrule
\begin{tabular}[c]{c} GPQA \\ ~\cite{rein2024gpqa} \end{tabular}
 & Domain-Specific & 546 & Text & Manual & N/A & Human & \begin{tabular}[c]{c}Acc, ECErr, \\Post-hoc\\ agreement\end{tabular}\\
\midrule
\begin{tabular}[c]{c} ScIRGen-Geo \\ ~\cite{lin2025scirgen} \end{tabular}
 & Domain-Specific & ~61k & Text & LLM, Manual & Static & Rules & Entailment Model Acc, Recall@k, MRR@100, ROUGE-L\\
\midrule
\begin{tabular}[c]{c} OlympiadBench \\ ~\cite{he2024olympiadbench} \end{tabular}
 & Domain-Specific & 8476 & Text, Image & \begin{tabular}[c]{c}Rules,\\LLM,\\ Manual\end{tabular} & Static & Rules, Human & Micro-average Acc, Acc \\
\midrule
\begin{tabular}[c]{c} DeepShop \\ ~\cite{lyu2025deepshop} \end{tabular}
 & Domain-Specific & 600 & Text, Image & LLM, Manual & Dynamic & \begin{tabular}[c]{c}Rules,\\LLM,\\Human\end{tabular} & Fine-grained SuccRate, Holistic Task SuccRate \\
\midrule
\begin{tabular}[c]{c} USACO \\ ~\cite{shi2024can} \end{tabular}
 & Domain-Specific & 307 & Text & \begin{tabular}[c]{c}Rules,\\LLM,\\Manual\end{tabular} & Static & Rules & Execution SuccRate, Pass@1 Acc \\
\midrule
\begin{tabular}[c]{c} GAIA \\ ~\cite{mialon2023gaia} \end{tabular}
 & Domain-Specific & 466 & Text, Image & Manual & Dynamic & Rules & Acc \\
\midrule
\begin{tabular}[c]{c} HLE \\ ~\cite{phan2025humanity} \end{tabular}
 & Domain-Specific & 2500 & Text, Image & LLM, Manual & Static & LLM & Acc, RMS \\
\midrule
\begin{tabular}[c]{c} HERB \\ ~\cite{choubey2025benchmarkingdeepsearchheterogeneous} \end{tabular}
 & Domain-Specific & 40704 & Text & LLM, Manual & Static & Rules, LLM & F1, Prec, Rec, AvgPerformScore, Likert Scale Rating \\
\midrule
\begin{tabular}[c]{c} RAGChecker \\ ~\cite{ru2024ragchecker} \end{tabular}
 & Others & 4162 & Text & LLM, Manual & Static & LLM & Prec, Rec, F1 \\
\midrule
\begin{tabular}[c]{c} ToolQA \\ ~\cite{zhuang2023toolqa} \end{tabular}
 & Others & 1530 & Text, Table & \begin{tabular}[c]{c}Rules,\\LLM,\\Manual\end{tabular} & Static & Rules & EM \\
\midrule
\begin{tabular}[c]{c} KILT \\ ~\cite{petroni2020kilt} \end{tabular}
 & Others & ~3.2M & Text & Rules, Manual & Static & Rules & Acc, EM, F1, ROUGE-L, R-Prec, Recall@k, KILT scores \\
\midrule
\begin{tabular}[c]{c} CONFLICTS \\ ~\cite{cattan2025dragged} \end{tabular}
 & Others & 458 & Text & Manual & Dynamic & LLM & \begin{tabular}[c]{c} Fact-Ground, \\AnsRec,\\ExpBeh\end{tabular} \\
\midrule
\begin{tabular}[c]{c} Search Arena \\ ~\cite{miroyan2025search} \end{tabular}
 & Others & 36721 & Text & Manual & Dynamic & Human & WinRate, Bradley-Terry model coefficients \\
\midrule
\begin{tabular}[c]{c} WebWalkerQA \\ ~\cite{wu2025webwalker} \end{tabular}
 & Others & 680 & Text & LLM, Manual & Dynamic & LLM, Rules & Acc, Action count \\
\midrule
\begin{tabular}[c]{c} IIRC \\ ~\cite{ferguson2020iirc} \end{tabular}
 & Others & 13441 & Text & Manual & Dynamic & Rules & EM, F1 \\
\midrule
\begin{tabular}[c]{c} Instruct2DS \\ ~\cite{ma2025autodata} \end{tabular}
 & Others & 234 & Text & Manual & Dynamic & Rules & \begin{tabular}[c]{c}F1, Prec, Rec, \\Exe-Prec, \\Exe-Rec\end{tabular} \\
\midrule
\begin{tabular}[c]{c} DRComparator \\ ~\cite{chandrahasan2025deep} \end{tabular}
 & Others & / & Text & Manual & Dynamic & Human & BT Score \\

\bottomrule
\end{longtable}
\normalsize
\twocolumn

\end{document}